\documentclass{openjournal}

\usepackage[utf8]{inputenc}
\usepackage[english]{babel}

\usepackage{amsmath}
\usepackage{amssymb}
\usepackage{bm}
\usepackage{xcolor}
\usepackage{graphicx}
\DeclareGraphicsExtensions{.bmp,.png,.jpg,.pdf}
\graphicspath{{./figs/}}

\usepackage{xcolor}
\usepackage{color,colortbl}
\definecolor{linkcolor}{rgb}{0.0,0.3,0.5}

\usepackage{textgreek}
\usepackage{verbatim}
\usepackage[normalem]{ulem}
\usepackage{soul}

\usepackage{tensind}
\tensordelimiter{?}

\usepackage{orcidlink}

\usepackage{hyperref}
\hypersetup{
    unicode=true,
    colorlinks=true,
    linkcolor=linkcolor,
    citecolor=linkcolor,
    filecolor=linkcolor,
    urlcolor=linkcolor,
}
\usepackage[capitalise]{cleveref}

\DeclareMathAlphabet{\mathup}{OT1}{\familydefault}{m}{n}

\newcommand{\be}{\begin{equation}} 
\newcommand{\ee}{\end{equation}}

\def\cdl#1{\textcolor{teal}{\tt [CDL: #1]}}

\def\ett#1{\textcolor{black}{#1}}

\begin{document}
\title{The road towards precision measurements of $H_0$ with bright sirens in the Einstein Telescope era}

\author{Chiara De Leo\orcidlink{0009-0009-1175-213X}}
\email{chiara.deleo@uniroma1.it}
\affiliation{Physics Department, Sapienza University of Rome, P.le A. Moro 5, 00185 Roma, Italy}
\affiliation{INAF - Osservatorio Astronomico di Roma, via Frascati 33, 00078 Monte Porzio Catone, Italy}
\affiliation{Istituto Nazionale di Fisica Nucleare (INFN), Sezione di Roma, P.le A. Moro 5, I-00185, Roma, Italy}
\affiliation{Physics Department, Tor Vergata University of Rome, Via della Ricerca Scientifica 1, Roma, 00133, Italy}

\author{Elsa M. Teixeira\orcidlink{0000-0001-7417-0780}\hspace{-0.6em}}
\email{elsa.teixeira@umontpellier.fr}

\author{\hspace{0.5em}Vivian Poulin\orcidlink{0000-0002-9117-5257}}
\email{vivian.poulin@umontpellier.fr}

\affiliation{
Laboratoire Univers \& Particules de Montpellier,\\
CNRS \& Université de Montpellier (UMR-5299), 34095 Montpellier, France
}

\def\cdl#1{\textcolor{black}{#1}}

\begin{abstract}
Gravitational-wave standard sirens provide an independent probe of cosmic expansion since their luminosity distances are inferred directly from the gravitational-wave signal and, for bright sirens (BSs), the source redshifts are obtained through the identification of electromagnetic counterparts. In this work, we forecast the constraining power of future bright siren catalogues on the Hubble constant using simulated binary-neutron-star mergers detected by the Einstein Telescope with associated electromagnetic counterparts.
We construct mock catalogues with different numbers of events and redshift distributions, with GW170817 as a reference BS, and analyse the resulting constraints within a flat $\Lambda$CDM cosmology. We find that, when bright sirens are used as a standalone probe, most of the information on $H_0$ is provided by low-redshift events, with the improvement in precision saturating for sources above $z \sim 1$; beyond that redshift, the constraining power of BSs is increasingly limited by the degeneracy with $\Omega_m$, which prevents further gains in precision on $H_0$. In this case, approximately $90$ low-redshift BSs are required to reach $\sigma_{H_0} \sim 1\,{\rm km\,s^{-1}\,Mpc^{-1}}$, while about $45$ are sufficient for $\sigma_{H_0} \sim 2\,{\rm km\,s^{-1}\,Mpc^{-1}}$. \cdl{When external BAO information is included to reduce the $H_0-\Omega_m$ degeneracy, intermediate-redshift sirens become more informative and the required number of events decreases substantially. We find that approximately 15 BSs are sufficient to reach $\sigma_{H_0}\sim2\,{\rm km\,s^{-1}\,Mpc^{-1}}$ for a typical catalogue realisation, while $\sigma_{H_0}\sim1\,{\rm km\,s^{-1}\,Mpc^{-1}}$ requires about 46 BSs, and can be reached with as few as about 20 BSs for a favourable redshift distribution of the events.} These results highlight the importance of both electromagnetic counterpart identification and complementary background probes in making BSs a competitive, distance-ladder-independent test of the Hubble tension.
\end{abstract}


\maketitle

\section{Introduction}\label{sec:intro}

The standard cosmological model, the $\Lambda$CDM model, posits the existence of a dark sector needed to explain current observations. Here, $\Lambda$ is the cosmological constant, a constant dark energy source to explain the late-time phase of accelerated expansion. On the other hand, cold dark matter (CDM) is introduced to provide a mechanism for the formation and dynamics of structures in the Universe. 
The current expansion rate of the Universe is parametrised by the Hubble constant $H_0=H(z=0)$, which is the normalisation constant of the background expansion rate:

\begin{equation}
    H(z) = H_0 \sqrt{\Omega_{\rm r,0} a^{-4} + \Omega_{\rm m,0} a^{-3} + \Omega_{\Lambda}}\, ,
    \label{eq:hz}
\end{equation}
in a spatially flat FLRW background, where $a$ is the scale factor of the Universe and $\Omega_{\rm r,0}$, $\Omega_{\rm m,0}$, and $\Omega_{\Lambda}$ are the present fractional densities of radiation, matter (CDM and baryons), and the cosmological constant. 
A direct probe of the expansion history of the Universe is the luminosity distance, $d_L$, which for a source at redshift $z$ can be estimated from

\begin{equation}
d_L(z) = c(1+z)\int_0^z \frac{{\rm d}z'}{H(z')} \, .
\label{eq:dl_flrw_bs}
\end{equation}

The Hubble constant sets the absolute scale of cosmological distances. Current determinations of $H_0$ show a significant discrepancy between values obtained from the late-Universe distance ladder and those inferred from early-Universe observations within the $\Lambda$CDM model. In particular, the S$H_0$ES distance-ladder analysis, based on Cepheid-calibrated Type Ia supernovae, gives $H_0 = 73.04 \pm 1.04 \, \mathrm{km\,s^{-1}\,Mpc^{-1}}$ \citep{SH0ES_H0}. On the other hand, observations of the cosmic microwave background (CMB) from the \textit{Planck} satellite, when interpreted assuming $\Lambda$CDM, give $H_0 = 67.4 \pm 0.5 \, \mathrm{km\,s^{-1}\,Mpc^{-1}}$ \citep{Planck2018}. This mismatch is commonly referred to as the Hubble tension and represents one of the most pressing open problems in modern cosmology. 

Several explanations have been proposed, depending on whether the tension is interpreted as the result of unaccounted observational systematics, as a hint of limitations in the assumptions made under $\Lambda$CDM or as a possible indication of physics beyond the standard model. On the observational side, the discrepancy could be related to residual systematics in the local distance ladder, for example in the calibration of Cepheid variables, in the standardisation of Type Ia supernovae, or in the connection between the different rungs of the ladder \citep{Freedman_2021,Efstathiou_2021}. In fact, the recent H$_0$ Distance Network consensus finds that the local determination is robust across first-rank distance indicators, with a covariance-weighted baseline value
$H_0 = 73.50 \pm 0.81\,{\rm km\,s^{-1}\,Mpc^{-1}}$; this makes a simple systematic error associated with a single calibrator, distance indicator, or ladder rung increasingly difficult to isolate as the sole origin of the tension \citep{H0DN:2025lyy}. On the theoretical side, if the tension is not due to systematics, it may suggest that some assumptions of the standard cosmological model need to be modified. Proposed solutions can be broadly divided into early-time mechanisms, which affect the pre-recombination Universe and reduce the sound horizon, and late-time mechanisms, which modify the expansion history at low redshift \citep{Verde_2019,DiValentino_2021,Poulin_2023}. 
In this context, independent probes of the distance--redshift relation are essential to test whether the discrepancy is due to the calibration of electromagnetic distance indicators, to the assumptions of $\Lambda$CDM, or to a combination of both. 

Bright standard sirens (BSs) provide such an independent route, since they measure luminosity distances directly from the gravitational-wave (GW) signal and obtain redshift information from an electromagnetic (EM) counterpart. For compact binary mergers, the observed waveform encodes the luminosity distance in terms of the amplitude and phase evolution. If an EM counterpart for the event is identified, the redshift of the host galaxy can also be retrieved, producing an independent measurement of the luminosity distance. The first measured BS was from the GW170817 event, a binary neutron star (BNS) merger detected in gravitational waves and associated with the EM counterpart GRB 170817A and the host galaxy NGC 4993 \citep{GW170817, Hjorth_2017, Cowperthwaite_2017}. For a single event, the resulting constraint on $H_0$ is much broader than current electromagnetic (e.g. from supernovae) and CMB determinations, but GW170817 demonstrated the potential of bright siren cosmology \citep{Abbott2017H0}. Future third-generation detectors, such as the Einstein Telescope \citep{ET_science}, are expected to detect many more events over a broad redshift range, and a subset of these events may have detectable electromagnetic counterparts, allowing them to be used as BSs.

The potential of gravitational-wave standard sirens to measure $H_0$ has been extensively investigated in the literature. \ett{Early studies of nearby binary neutron star mergers with identified electromagnetic counterparts established the prospects of using joint GW and short-GRB observations as an independent probe of the expansion at low redshift \citep{Nissanke:2009kt,Nissanke:2013fka}. Later forecasts showed that samples of order $50$--$100$ bright sirens observed with the current Ligo--Virgo--Kagra (LVK) detector network could reach percent-level precision on $H_0$, potentially distinguishing between the local and CMB-inferred values of $H_0$ \citep{Chen:2017rfc,Feeney:2018mkj,Palmese:2019ehe,Mortlock:2018azx}. } Subsequent studies extended these analyses to third-generation observatories such as the Einstein Telescope (ET), considering larger catalogues spanning a broader range of redshifts \citep{Califano:2022cmo,Cozzumbo:2024vxw,deSouza:2021xtg,Han:2023exn,Sathyaprakash:2009xt,Taylor:2012db}. The complementarity between standard sirens and electromagnetic probes, including BAO, has also been explored as a way of breaking degeneracies in parameters describing the background expansion, such as $H_0$ and $\Omega_m$~\citep{Chang_2019, Zhang_2019,Zhao:2010sz,Cai:2016sby} and to constrain extended models \citep{Teixeira:2023zjt,Giare:2024syw,DeLeo:2025lmx,Graziotti:2026qzl,Califano:2022syd}. 

The importance of the redshift distribution for BSs has also been the focus of previous work. At sufficiently low redshift, the luminosity distance in Eq.~\ref{eq:dl_flrw_bs} is dominated by the leading-order relation $d_L(z) \approx c z / H_0$, so nearby events provide a comparatively direct estimate of $H_0$. At higher redshift, the distance-redshift relation becomes increasingly sensitive to the matter density and to the assumed background history. In the context of third-generation GWs, most of the information on $H_0$ has been found to be supplied by local events and that sources above $z\sim 1$ provide only limited additional improvement, as shown for dark sirens \citep{Muttoni:2023prw}. These results suggest that the constraining power of standard sirens does not depend solely on the size of the catalogue, but also on its redshift distribution, distance uncertainties, EM counterpart selection function and the availability of complementary cosmological information. \ett{In addition, requiring an EM counterpart introduces a non-trivial selection in source inclination, whose consistent treatment is important for an unbiased inference of $H_0$ \citep{Mancarella:2024qle}.}

Building on these studies, our aim is not to establish the general low-redshift sensitivity of standard siren measurements, but to quantify it for a population of ET binary neutron star detections with identified EM counterparts. We investigate how different portions of the resulting BS catalogue contribute to the marginalised constraint on $H_0$ and how this redshift dependence changes when the sirens are combined with mock future BAO measurements from SKA Observatory (SKAO). We then estimate the number and redshift distribution of identified bright sirens required for the resulting measurement of $H_0$ to reach a precision of $\sigma_{H_0} \sim 1\,{\rm km\,s^{-1}\,Mpc^{-1}}$ and $\sigma_{H_0} \sim 2\,{\rm km\,s^{-1}\,Mpc^{-1}}$, respectively.

\begin{figure}[h!]
\centering
\includegraphics[width=0.75\linewidth]{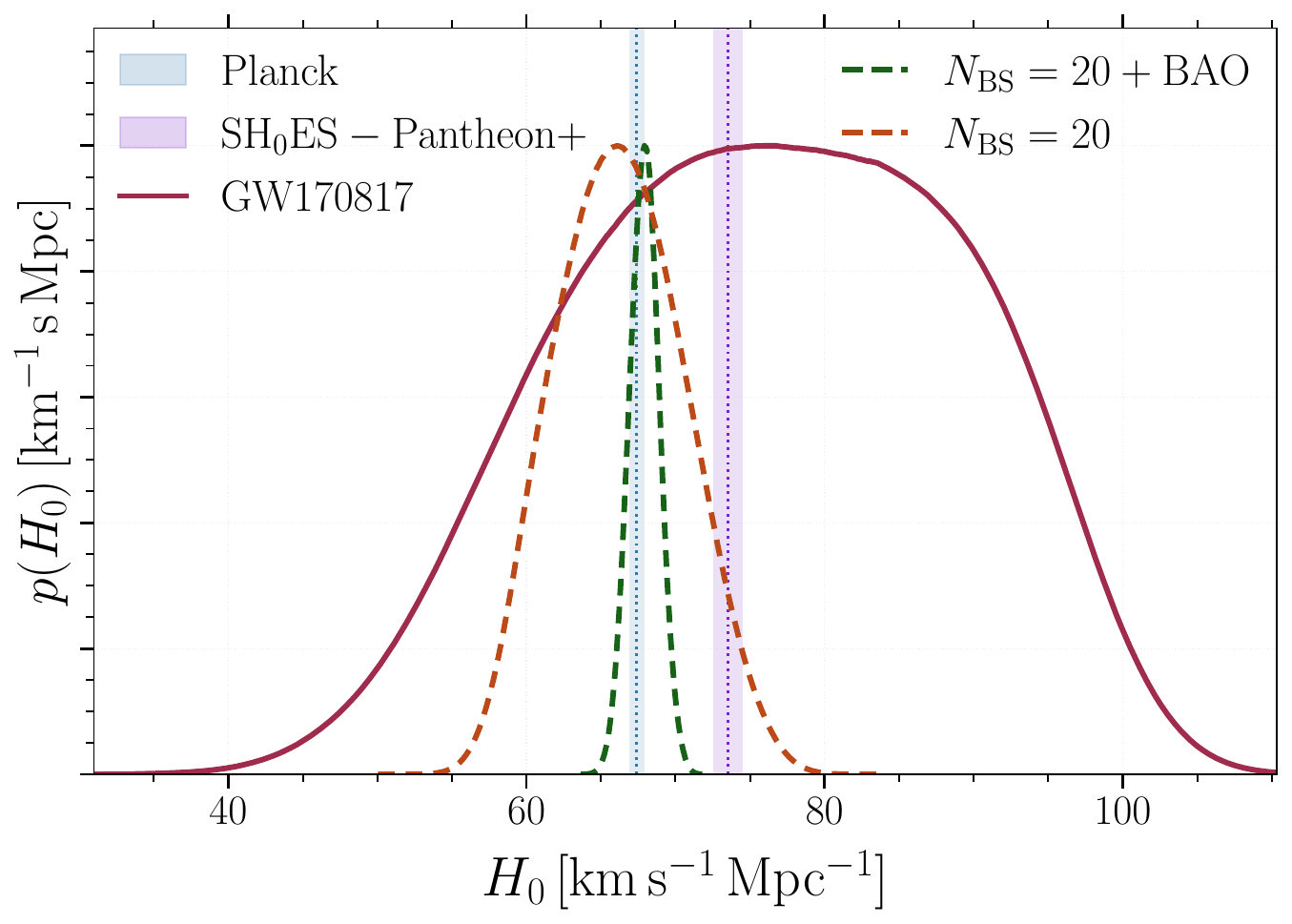}
\caption{Posterior probability distribution of $H_0$ inferred from GW170817, compared with the S$H_0$ES-calibrated Pantheon$+$ $H_0$ (purple) measurement, the value constrained from \textit{Planck} under a $\Lambda$CDM cosmology (blue) and the posterior obtained using \cdl{only $19$ mock bright-siren events and GW170817} (dashed orange) and analysed with external BAO information (dashed green). The mocks are generated following the prescriptions described in Sections~\ref{sec:data_bs} and \ref{sec:data_BAO}. }
\label{fig:GW170817_BAO_H0}
\end{figure}

Our main result is presented in Fig.~\ref{fig:GW170817_BAO_H0}, where we show the posterior obtained from the only BS event detected to date, GW170817, compared with one of the mock BS catalogues analysed alone, represented by the dashed orange curve, and the corresponding posterior obtained after including external BAO information, represented by the dashed green curve. %\sout{\cdl{We note that our treatment of GW170817 uses its luminosity-distance measurement and uncertainty, rather than the full Bayesian parameter-estimation used by LVK as explained in \citet{Abbott2017H0}.}} 
\ett{We note that our treatment of GW170817 approximates the marginalised luminosity-distance likelihood as a Gaussian centred at $d_L=40\,{\rm Mpc}$, with uncertainty $\sigma_{d_L}=11\,{\rm Mpc}$, obtained by symmetrising the measured $68\%$ interval $40^{+8}_{-14}\,{\rm Mpc}$ at $z=0.0098$. Therefore, this simplified treatment does not reproduce the full non-Gaussian distance-inclination posterior or propagate the uncertainty in the peculiar velocity of NGC~4993, as explained in \citet{Abbott2017H0,LIGOScientific:2026uyd}.} We find that about 20 BS events detected by ET in combination with a constraint on $\Omega_m$ from BAO data can reach $\sigma_{H_0}\sim1\,{\rm km\,s^{-1}\,Mpc^{-1}}$. \cdl{ By repeating the analysis over 1000 realisations of the catalogue (Sec.~\ref{sec:realisations}), we show that this is a favourable case: a typical realisation requires about 46 events to reach the same precision, while about 15 are sufficient for $\sigma_{H_0}\sim2\,{\rm km\,s^{-1}\,Mpc^{-1}}$.}

This paper is structured as follows. We first introduce cosmology with BSs in Section~\ref{sec:bs}. We then describe the observables and datasets used in the analysis in Section~\ref{sec:data}. In Section~\ref{sec:methodology}, we present the methodology adopted to obtain the results, which are discussed in Section~\ref{sec:results} and Section~\ref{sec:realisations}. Finally, we draw our conclusions in Section~\ref{sec:conclusions}.

\section{Bright Sirens}
\label{sec:bs}

Gravitational waves can be described as small perturbations $h_{\mu\nu}$ of the space-time metric. In a flat FLRW expanding Universe, tensor perturbations propagate according to
\begin{equation}
h''_{+,\times}(\tau,k)
+ 2\mathcal{H} h'_{+,\times}(\tau,k)
+ k^2 h_{+,\times}(\tau,k) = 0 \, ,
\label{eq:gw_propagation_bs}
\end{equation}
where primes denote derivatives with respect to conformal time $\tau$, $k$ is the comoving wavenumber, $\mathcal{H}=a'/a$ is the conformal Hubble parameter, and $h_+$ and $h_\times$ are the two GW polarizations. Here, the friction term describes the damping of the GW amplitude due to the cosmic expansion.

A variety of astrophysical and cosmological mechanisms are expected to generate GWs, including compact binary coalescences (CBCs), core-collapse supernovae, rotating non-axisymmetric neutron stars, stochastic backgrounds of astrophysical or primordial origin, and possible signals from the early Universe. To date, however, all confirmed direct GW detections have been associated with CBCs involving binary black holes, binary neutron stars, and neutron-star--black-hole systems \citep{Abbott_2016, GW170817, GWTC4, GWTC5}. These sources are particularly relevant for cosmology because their waveform provides a direct and independent measurement of the luminosity distance. Unlike EM probes, which rely on the calibration of standard candles or standard rulers, CBCs are self-calibrating sources: the distance information is encoded in the amplitude and phase evolution of the signal. For this reason, they are commonly referred to as \textit{standard sirens} \citep{Schutz_1986, Holz_2005}.

For a compact binary inspiral, at leading order and in the frequency domain, the strain amplitude scales as
\begin{equation}
\tilde{h}(f_{\rm d}) \propto
\frac{\mathcal{M}_{\rm d}^{5/6}}{d_L}\,
f_{\rm d}^{-7/6}
e^{i\Psi(f_{\rm d},\mathcal{M}_{\rm d})}\, ,
\label{eq:strain_amplitude}
\end{equation}
where $f_{\rm d}$ is the detector-frame frequency, $\Psi$ is the GW phase, and $\mathcal{M}_{\rm d}$ is the detector-frame chirp mass. The detector-frame quantities are related to the source-frame ones by
\begin{equation}
f_{\rm d}=\frac{f_{\rm s}}{1+z},
\qquad
\mathcal{M}_{\rm d}=(1+z)\mathcal{M}_{\rm s}\, .
\label{eq:redshifted_quantities_bs}
\end{equation}
Therefore, the GW signal directly constrains the luminosity distance, $d_L$, and the redshifted chirp mass, $\mathcal{M}_{\rm d}$. However, because the source redshift cannot be determined from the GW signal alone, an independent redshift measurement is essential for cosmological applications \citep{Maggiore_2018, Palmese_2025, pierra_2025}.

When an electromagnetic counterpart, such as a short gamma-ray burst or a kilonova, is identified, the host galaxy can be associated with the GW event and its redshift can be measured directly. These events are known as \textit{bright sirens}. They differ from dark sirens, for which no EM counterpart is observed and the redshift must instead be inferred statistically \citep{Palmese:2021mjm}.

In this work, we focus on bright sirens. They represent the cleanest way of using GWs for cosmology, since each event provides both a GW luminosity distance and an EM redshift. The cosmological information is then extracted by comparing the measured $d_L$--$z$ relation with the theoretical prediction, according to Eq.~\ref{eq:dl_flrw_bs}.
At low redshift, this relation is mainly sensitive to the Hubble constant $H_0$. For this reason, BSs are particularly relevant in the context of the Hubble tension, since they offer an independent measurement of $H_0$ that does not rely either on the cosmic microwave background or on the local distance ladder.

\section{Dataset}
\label{sec:data}
\subsection{Current bright-siren data} \label{sec:current_bs}
As discussed in Sections~\ref{sec:intro} and~\ref{sec:bs}, bright sirens provide an independent measurement of the cosmic expansion and therefore represent a valuable probe of the Hubble tension. 

To date, GW170817 is the only gravitational-wave event with an identified electromagnetic counterpart that has provided a bright-siren measurement of $H_0$ \citep{GW170817,Hotokezaka:2018dfi,Palmese:2023beh}. It was the first binary neutron-star inspiral detected by Advanced LIGO and Advanced Virgo, with a combined signal-to-noise ratio of $32.4$ and a sky localisation of $28\,\mathrm{deg}^2$ at $90\%$ probability. The GW signal provided a direct measurement of the luminosity distance, $d_L = 40^{+8}_{-14}\,\mathrm{Mpc}$ \citep{GW170817_dL}, while the subsequent identification of electromagnetic counterparts, including GRB~170817A and the host galaxy NGC~4993, supplied the redshift information, $z \simeq 0.0098$ \citep{Hjorth_2017,Howlett:2019mdh}. \ett{In parallel, increasingly large samples of dark sirens have been combined with galaxy catalogues to obtain independent constraints on $H_0$ \citep{Palmese:2021mjm,LIGOScientific:2026uyd}.}

In this work, we analyse GW170817 using the BS likelihood implemented in the \texttt{CANDI}\footnote{\href{https://github.com/chiaradeleo1/CANDI}{https://github.com/chiaradeleo1/CANDI}} code \citep{DeLeo:2025rhy}, which combines the measured luminosity distance and source redshift to derive posterior constraints on the cosmological parameters. Although GW170817 established the potential of BSs as independent cosmological probes, its constraint on $H_0$ remains substantially broader than those obtained from electromagnetic observations. As shown in Fig.~\ref{fig:GW170817_BAO_H0}, the current BS measurement is therefore not yet competitive with either the distance-ladder determination or the CMB-inferred constraint.

This naturally raises the question of how many additional BSs would be required for a GW-based measurement of $H_0$ to become competitive with current electromagnetic constraints from S$H_0$ES-Pantheon$+$ and \textit{Planck}. Reaching this level of precision would make BSs a powerful independent tool for investigating whether the observed discrepancy arises from astrophysical systematics or assumptions about the cosmological model.

To address this question, in the following section we construct mock catalogues of future BS events and analyse them both independently and in combination with external mock BAO information. 
\subsection{The bright sirens catalogue} \label{sec:data_bs}

The main goal of this work is to assess how many BS events are required to constrain the Hubble constant $H_0$ with a precision competitive with current electromagnetic constraints, in particular the S$H_0$ES-calibrated Pantheon$+$ \citep{SH0ES_H0} and \textit{Planck} \citep{Planck2018}. To this end, we build a mock catalogue of BNS mergers with associated electromagnetic counterparts. The full pipeline used to generate the catalogue is publicly available in the code \texttt{CANDI}.
\noindent
Our simulations are based on the expected performance of the Einstein Telescope \citep{ET_science}, a forthcoming third-generation ground-based GW detector. We follow the specifications described in \citet{Branchesi_2023}, assuming a triangular configuration with $10 \, \mathrm{km}$ arms located in Sardinia, among the possible configurations.
We note that alternative design choices, such as adopting two L-shaped interferometers instead of a triangular layout, can change the detector sensitivity and the expected event rate \citep{Branchesi_2023}. These differences may therefore affect the number and redshift distribution of the detected BSs and, consequently, the quantitative forecasts presented in this work.

We generate $N=20000$ injected GW signals from BNS mergers in the redshift range $0.1<z<3.5$. This interval is chosen to reproduce the BNS detections expected for the Einstein Telescope \citep{Branchesi_2023}. We note, however, that the lower cut at $z=0.1$ excludes the very nearby population. Therefore, GW170817 will always lie below the redshift range of our simulated catalogues, thereby providing the lower-redshift reference point throughout our analysis. This is a limitation of the adopted ET-oriented mock construction, rather than an indication that bright sirens cannot occur at lower redshift. 

The source population is distributed according to a redshift-dependent merger rate, following \citet{Cutler_2009, Hogg_2020},
\begin{equation}
\label{eq:merger_rate}
R(z) =
\begin{cases}
1 + 2z\, , & z \leq 1\, , \\
\dfrac{3}{4}(5-z)\, , & 1 < z < 5\, , \\
0\, , & z \geq 5\, .
\end{cases}
\end{equation}
The corresponding redshift distribution is given by
\begin{equation}
P(z) \propto \frac{R(z)\,4\pi d_C^2(z)}{H(z)(1+z)}\, ,
\end{equation}
where $d_C(z)$ is the comoving distance and the factor $(1+z)^{-1}$ accounts for the conversion between source-frame and observer-frame time. The inclination angle $\iota$ is drawn assuming $\cos\iota$ uniformly distributed in the interval $[-1,1]$.
\noindent
The events are simulated using a modified version of the public code \texttt{darksirens}\footnote{\href{https://gitlab.com/matmartinelli/darksirens}{https://gitlab.com/matmartinelli/darksirens}} \citep{Martinelli_2022}. The code implements the merger rate in Eq.~\eqref{eq:merger_rate} and assumes a monochromatic neutron-star mass of $1.4\,M_\odot$.

For each event, the GW luminosity distance $d_L^{\rm GW}$ is computed using the theory module of \texttt{CANDI}. As fiducial cosmology, we adopt the best-fit values obtained from \textit{Planck}+BAO \citep{Planck2018} in $\Lambda$CDM. Specifically, we use
\begin{equation} \label{eq:fiducial}
H_0 = 67.66\,{\rm km\,s^{-1}Mpc^{-1}}\, ,
\qquad
\Omega_m = 0.3111\, .
\qquad
\end{equation}
The uncertainties on the GW luminosity distance and the detectability of the events are evaluated using the public code \texttt{GWFish}\footnote{\href{https://github.com/janosch314/GWFish}{https://github.com/janosch314/GWFish}} \citep{Dupletsa_2023}. We apply a signal-to-noise-ratio threshold and consider an event as detected if
${\rm SNR} > {\rm SNR}_{\rm th}=20,$
which corresponds to the most conservative detectability threshold considered for ET in \citet{Branchesi_2023}.

For the electromagnetic counterparts, we focus on the afterglow emission associated with short gamma-ray bursts (GRBs), while we do not include kilonova emission \citep{Metzger_2019,Han:2025fii}. The GRB emission is produced by a highly collimated relativistic jet, which generates a prompt gamma-ray signal followed by a broadband afterglow extending from X-rays to radio wavelengths. The afterglow can remain observable over long timescales, up to years after the merger. Kilonova emission, instead, arises from neutron-rich ejecta and is typically more isotropic, peaking in the optical or infrared bands depending on the ejecta composition.

To model the detectability of the EM counterpart, we follow the approach presented in \citet{DeLeo:2025rhy}. In that work, the Vera Rubin Observatory/LSST sensitivity \citep{Bianco2022} was adopted as a reference for EM detectability, and the afterglow emission was modelled using the code \texttt{afterglowpy}\footnote{\href{https://github.com/geoffryan/afterglowpy}{https://github.com/geoffryan/afterglowpy}} \citep{Ryan_2020}. It was shown (see Figure 3 of \citet{DeLeo:2025rhy}) that, in terms of the final number of detectable counterparts, this procedure is approximately equivalent to applying a cut on the binary inclination angle. 
Following this result, we select BSs by imposing
$\iota < \iota_{\rm cut}=18^\circ$. \ett{We note that the fact that EM counterparts are predominantly selected at small inclination angles introduces a selection effect \citep{Mancarella:2024qle,Chen:2020dyt,Chen:2023dgw}. While a fully hierarchical treatment should account for this selection in the cosmological likelihood, in this work we take the inclination cut as a simplified forecast method calibrated to reproduce the expected results.}

To perform the analysis presented in Section~\ref{sec:results}, and in particular to study how the constraint on $H_0$ depends on both the number of BSs and their redshift distribution, we construct several sub-catalogues from the master catalogue, each containing a different number of events. The detailed properties of these sub-catalogues are discussed in the corresponding parts of Section~\ref{sec:results}.
\noindent
All sub-catalogues include GW170817, which represents the $N=1$ case. \cdl{We make this choice because GW170817 is currently the only observed bright siren, so any future analysis based on bright sirens will necessarily include it. Moreover, as shown by its posterior on $H_0$ in Fig.~\ref{fig:GW170817_BAO_H0}, the constraint from this single event is broad, so including it does not bias our results and leaves them essentially unchanged.} Larger catalogues are then built by adding $N-1$ events randomly selected from the master catalogue. The catalogues are constructed progressively, so that samples with larger $N$ contain the events included in the smaller samples plus additional randomly selected BSs. This procedure allows us to track how the uncertainty on $H_0$ evolves as the number of BS events increases. We emphasise that by including the observed GW170817 we are combining simulated measurements using the forecast ET sensitivity with a real LIGO-Virgo data point. This approach is statistically consistent since all the events are treated separately (diagonal covariance matrix) and we have considered a Gaussian distance uncertainty for GW170817.
Nevertheless, this means that the smallest catalogues can be affected by the particular noise realisation, while the larger catalogues are progressively dominated by the simulated ET population. \cdl{The impact of the specific realisation on our results is quantified in Sec.~\ref{sec:realisations}.}

\subsection{The baryon acoustic oscillations mock catalogue}
\label{sec:data_BAO}

Bright sirens alone cannot fully disentangle the effects of $H_0$ and $\Omega_m$ in the background cosmology. This can be understood from the fact that BSs constrain the expansion history through the luminosity distance, which depends on the Hubble function. In a flat $\Lambda$CDM background, neglecting the radiation contribution, Eq.~\ref{eq:hz} can be rewritten as
\begin{equation}
\label{eq:Hz}
H(z) = H_0 \sqrt{\Omega_m(1+z)^3 + (1-\Omega_m)} \, .
\end{equation}
Since $H(z)$ depends on both $H_0$ and $\Omega_m$, these parameters can be partially degenerate in a BS-only analysis. For this reason, we include external information on $\Omega_m$ from baryon acoustic oscillations (BAO).

The complementarity between standard sirens and BAO in reducing the $H_0$--$\Omega_m$ degeneracy has already been discussed in previous studies \citep{Chang_2019,Zhang_2019}. As we show in Section~\ref{sec:results}, adding BAO information not only tightens the final constraint on $H_0$, but also changes which part of the BS catalogue carries useful information on this parameter, making intermediate-redshift events more informative once the degeneracy with $\Omega_m$ is reduced.

BAO measurements constrain combinations of cosmological distances rescaled by the sound horizon at the drag epoch, $r_d$. In a flat Universe, the relevant distance measures are
\begin{equation}
\label{eq:bao_distances}
    d_M(z) = \frac{c}{H_0}\int_0^z \frac{{\rm d}z'}{E(z')}\, ,
    \qquad
    d_H(z) = \frac{c}{H(z)}\, ,
    \qquad
    d_V(z) = \left[z d_M^2(z)d_H(z)\right]^{1/3}\, ,
\end{equation}
where $E(z)=H(z)/H_0$. BAO catalogues provide measurements of $d_M/r_d$, $d_H/r_d$, and their combination through $d_V/r_d$.

Such measurements, for example from galaxy and Lyman-$\alpha$ surveys by SDSS \citep{eBOSS:2020fvk} and DESI \citep{DESI:2025zgx}, are routinely used for precision cosmology. In our forecast, however, we use a mock BAO catalogue rather than current measurements, since the bright siren sample considered here corresponds to a future ET-era dataset. We adopt an SKAO-like forecast as a representative future BAO probe, allowing us to quantify how complementary geometric information can reduce the $H_0-\Omega_m$ degeneracy without tying the result to a particular existing BAO data release. Following \citet{DeLeo:2025rhy}, we adopt the optimistic SKA2 configuration \citep{SKAO_2020} and assume the same fiducial cosmology used for the BS catalogue, described in Section~\ref{sec:data_bs}. The mock measurements cover the redshift range $z \in [0.2,2]$, divided into 18 bins with width $\Delta z = 0.1$.

Since BAO constrain distances only through their ratio with $r_d$, additional assumptions would be required to extract absolute distances. In this work, we avoid introducing external priors on $r_d$, such as those from BBN, in order to keep the analysis independent of early-Universe calibration assumptions. We therefore directly simulate measurements of $d_M/r_d$ and $d_H/r_d$, propagating the expected relative errors from \citet{Bull_2016}.

\section{Methodology}\label{sec:methodology}
The goal of this work is to assess the requirements for future BS detections to reach a precision on $H_0$ competitive with the S$H_0$ES-calibrated Pantheon$+$ \citep{SH0ES_H0} and \textit{Planck} \citep{Planck2018}. In this way, BSs could become an independent cosmological observable, not relying on the distance ladder, and thereby providing new insight into the Hubble tension.

To this end, we reconstruct the posterior distribution of $H_0$ using the different sub-catalogues described in Section~\ref{sec:data}. The inference is performed within a Bayesian framework \citep{bayesian}, and the comparison between the simulated data and theoretical predictions is carried out using the public Bayesian analysis code \texttt{Cobaya}\footnote{\href{https://github.com/CobayaSampler/cobaya}{https://github.com/CobayaSampler/cobaya}} \citep{Cobaya}. For this purpose, we construct external likelihood modules able to handle the datasets discussed in Section~\ref{sec:data}.

The mock BS and BAO datasets are analysed using dedicated likelihood modules implemented within the \texttt{CANDI} framework. Within the Bayesian approach, the posterior probability distribution of the cosmological parameters $\boldsymbol{\theta}$, given a data vector $\mathbf{d}$, is obtained through Bayes' theorem,
\begin{equation}
\mathcal{P}(\boldsymbol{\theta}\mid\mathbf{d})
=
\frac{
\mathcal{L}(\mathbf{d}\mid\boldsymbol{\theta})
\,\pi(\boldsymbol{\theta})
}{
\mathcal{Z}
}\, ,
\label{eq:bayes_theorem}
\end{equation}
where $\mathcal{L}(\mathbf{d}\mid\boldsymbol{\theta})$ is the likelihood function, $\pi(\boldsymbol{\theta})$ denotes the prior distribution of the cosmological parameters, and $\mathcal{Z}$ is the Bayesian evidence. Assuming Gaussian-distributed measurements, each likelihood can be expressed as
\begin{equation}
-2\ln\mathcal{L}
=
\left[
\mathbf{d}-\mathbf{t}(\boldsymbol{\theta})
\right]^{\mathrm{T}}
\mathbf{C}^{-1}
\left[
\mathbf{d}-\mathbf{t}(\boldsymbol{\theta})
\right]\, ,
\label{eq:gaussian_likelihood}
\end{equation}
where $\mathbf{t}(\boldsymbol{\theta})$ is the theoretical prediction corresponding to the sampled cosmological parameters and $\mathbf{C}$ is the covariance matrix of the data. 
The data vector used in the likelihood contains the mock bright siren luminosity distances and BAO distance ratios,
$\mathbf{d}\equiv\{d_L(z), d_M(z)/r_d, d_H(z)/r_d\}$, constructed as described above.
For the bright siren catalogue, we assume that the luminosity distance measurements of distinct events are statistically independent, meaning that $\mathbf{C}$ is taken to be diagonal. This approximation neglects possible correlated systematic uncertainties, such as a common detector-calibration error. For the BAO mock, we neglect correlations between different redshift bins, but include the forecast covariance between the transverse and radial measurements, $d_M/r_d$ and $d_H/r_d$ within each bin.

For the theoretical modelling, we compute the standard cosmological quantities using \texttt{CAMB}\footnote{\href{https://camb.readthedocs.io/}{https://camb.readthedocs.io/}} \citep{Lewis:1999bs,2012JCAP...04..027H}, interfaced with a theory module that computes the background quantities. The full setup, made publicly available as \texttt{CANDI}\footnote{\href{https://github.com/chiaradeleo1/CANDI}{https://github.com/chiaradeleo1/CANDI}}, allows \texttt{Cobaya} to perform Markov Chain Monte Carlo (MCMC) sampling and reconstruct the posterior distribution of the cosmological parameters \citep{Lewis:2002ah,Lewis:2013hha}.

In all analyses, $H_0$ and $\Omega_m$ are treated as free cosmological parameters, with uniform priors over the intervals $H_0 \in [20,100]\,\mathrm{km\,s^{-1}\,Mpc^{-1}}$ and $\Omega_m \in [0.1,0.9]$. When BAO measurements are included, we do not impose any BBN information and instead sample directly over the sound horizon at the drag epoch, $r_d$ with a uniform prior $r _d\in [100,200]\, \text{Mpc}$.

\section{Results}
\label{sec:results}

In Section~\ref{sec:current_bs}, we showed that GW170817 provides the first proof that bright sirens can be used as an independent cosmological probe. However, as illustrated in Fig.~\ref{fig:GW170817_BAO_H0}, the current constraint on $H_0$ remains significantly broader than those obtained from electromagnetic observations, highlighting the need for substantially larger BS catalogues. In this section, we investigate how many BS detections will be required to achieve a precision of approximately $1\,{\rm km\,s^{-1}\,Mpc^{-1}}$ and $2\,{\rm km\,s^{-1}\,Mpc^{-1}}$ on $H_0$, comparable to that of current leading cosmological probes. We also explore whether the constraining power of BSs depends solely on the number of detected events or whether other factors, such as the redshift distribution of the sources and the inclusion of complementary cosmological observations, play a significant role in improving the measurement.

\subsection{Impact of the redshift of events}
\label{sec:results_redshift_number}

We first investigate how the constraint on $H_0$ depends on the number of BSs and on their redshift distribution. As described in Section~\ref{sec:data_bs}, we construct a master mock catalogue and use it to generate several sub-catalogues with different numbers of events $N$, always including GW170817.

\begin{figure}
\centering

\begin{minipage}{0.49\linewidth}
\centering
\includegraphics[width=\linewidth]{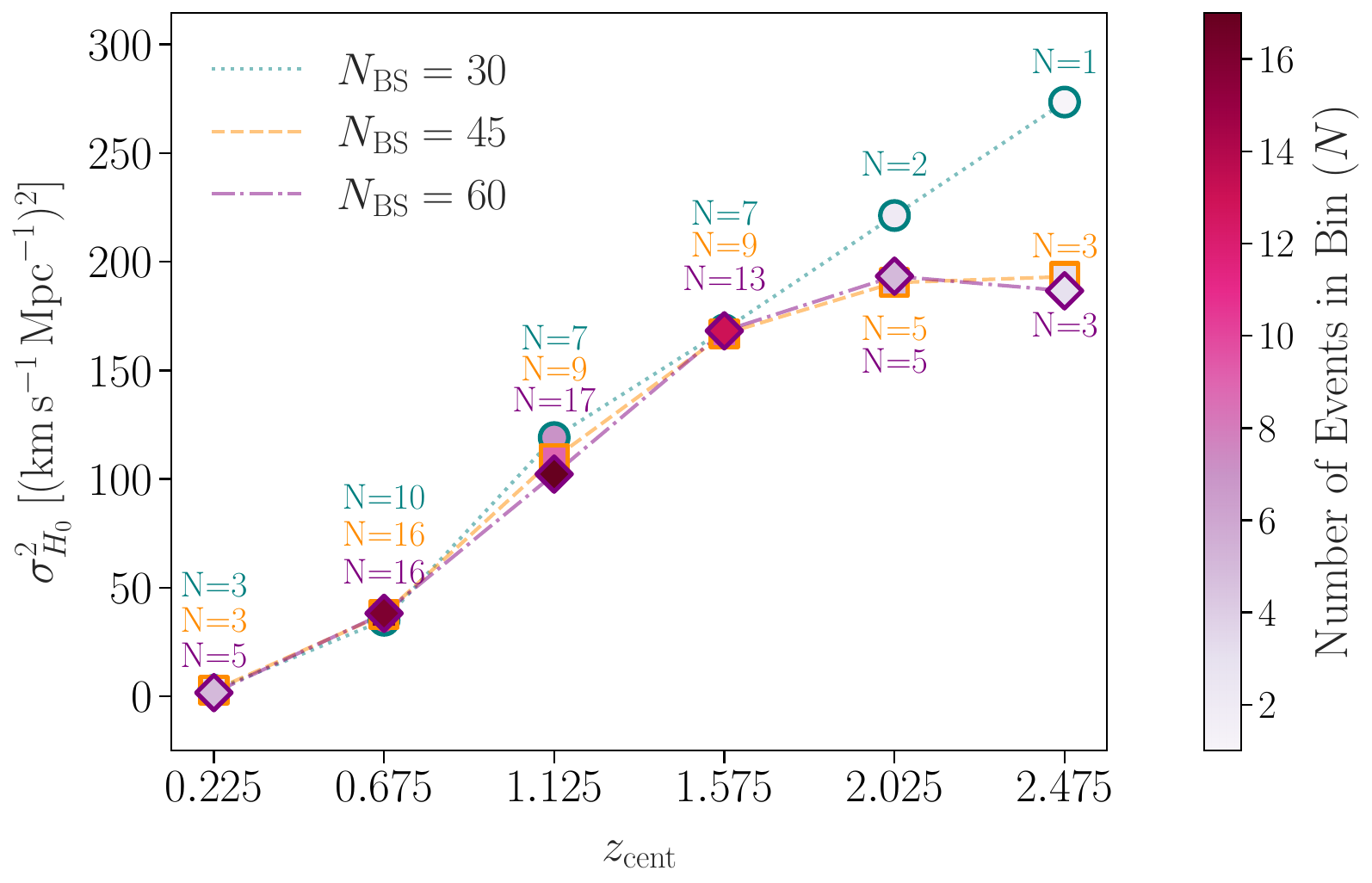}\
\textbf{(a)}
\label{fig:variance_zbins}
\end{minipage}
\hfill
\begin{minipage}{0.49\linewidth}
\centering
\includegraphics[width=\linewidth]{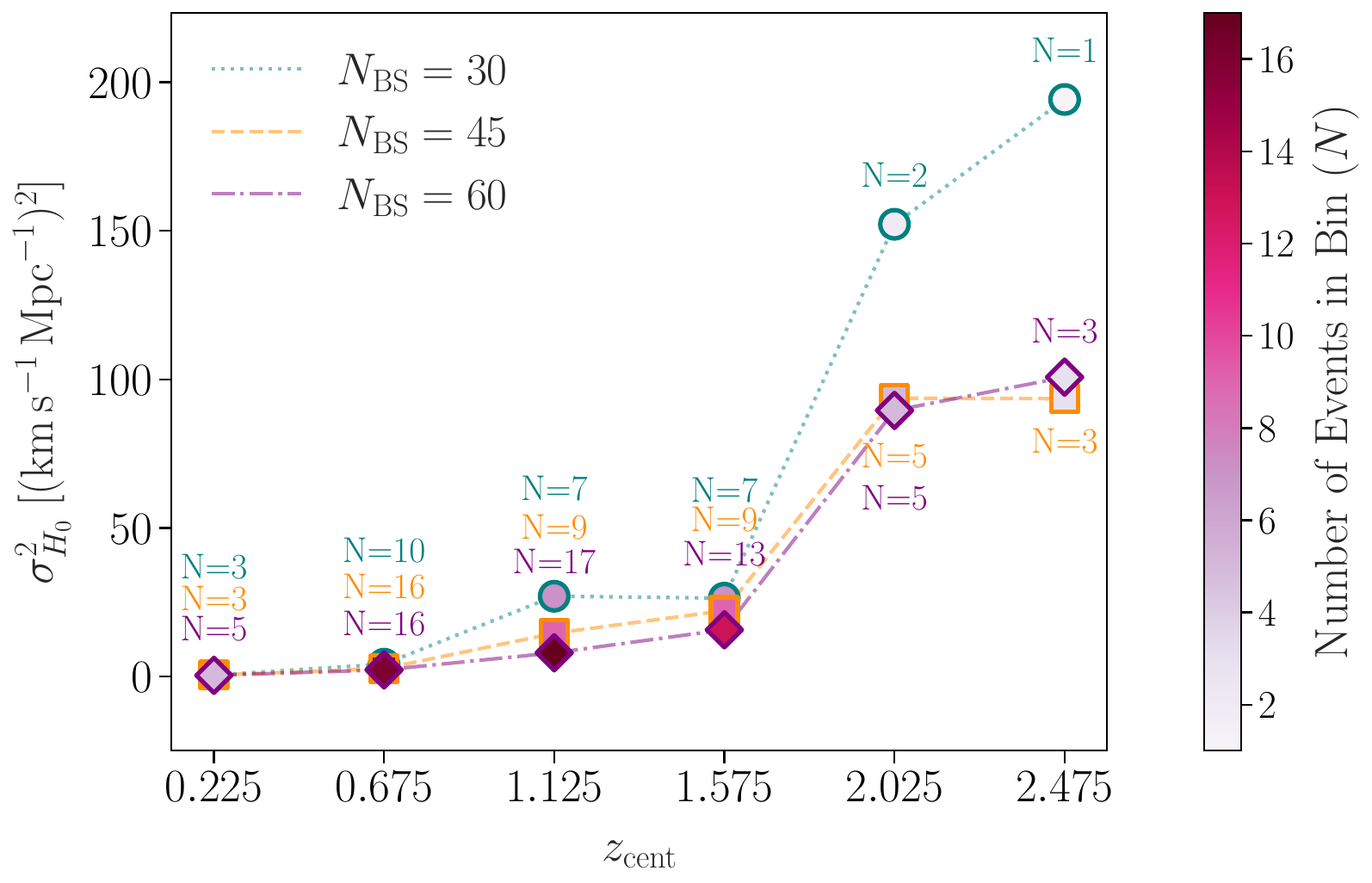}\
\textbf{(b)}
\label{fig:variance_zbins_omegam}
\end{minipage}

\caption{Variance of the inferred $H_0$ posterior as a function of redshift bin for three mock bright siren catalogues containing $N=30$, $N=45$, and $N=60$ events. The events are divided into six equally spaced redshift bins centred at $z_{\rm cent}$ and the labels indicate the number of events in each bin. Panel (a) shows the results obtained when only BS data are included, while panel (b) shows the corresponding results when complementary BAO information is included.}
\label{fig:variance_zbins_comparison}
\end{figure}

To assess the role of redshift, for each value of $N$ we divide the events into six equally spaced redshift bins between $z_{\rm min}$ and $z_{\rm max}$. These limits are defined from the largest sub-catalogue considered in this part of the analysis, containing $N=60$ events. Each bin is then analysed separately by sampling for $H_0$ and $\Omega_m$ only, and computing the variance of the marginalised $H_0$ posterior, $\sigma^2_{H_0}$. The results are shown in Fig.~\ref{fig:variance_zbins_comparison} for three representative catalogues with $N=30$, $N=45$, and $N=60$ events.

The binned analysis reveals a clear dependence of the marginalised $H_0$ constraint on redshift. 
As shown in panel (a) of Fig.~\ref{fig:variance_zbins_comparison}, the lowest redshift bin, centred around $z_{\rm cent}\simeq 0.23$, consistently yields the smallest variance on the $H_0$ posterior. The variance increases as we move towards higher redshift for all three catalogue sizes, indicating that events in the lower-redshift part of the catalogue provide a more direct constraint on $H_0$. This trend cannot be attributed simply to the number of events in each bin since, for example, the third bin contains 7, 9, and 17 events in the $N=30, 45$ and $ 60$ catalogues, respectively, but the resulting variances remain of comparable magnitude, suggesting that adding events in this redshift range produces only a limited improvement in the marginalised constraint on $H_0$.

A further feature emerging from this analysis is that, at high redshift, the variance tends to approach a plateau. This behaviour is visible for the larger catalogues, while it is less evident for the $N=30$ case, where the number of events in the last two bins is too small. These results indicate that the most informative events for constraining $H_0$ are those at $z\lesssim 1$, where the variance is smaller and the degeneracy with $\Omega_m$ is less severe. 
 
This behaviour is qualitatively expected: as already discussed in the introduction, the luminosity distance becomes increasingly sensitive to $\Omega_m$ at higher redshift, while its statistical uncertainty also grows with distance, both of which degrade the amount of information a given event carries on $H_0$ \citep{deSouza:2021xtg,Mpetha:2024xiu,Vaskonen:2026ubd,Canevarolo:2023dkh}. The main result of this analysis is to make this expectation quantitative for our BS population: rather than declining smoothly with redshift, the constraining power on $H_0$ is concentrated at $z\lesssim 1$, beyond which additional events add comparatively little to the marginalised constraint.

\begin{figure}
\centering

\begin{minipage}{0.49\linewidth}
\centering
\includegraphics[width=\linewidth]{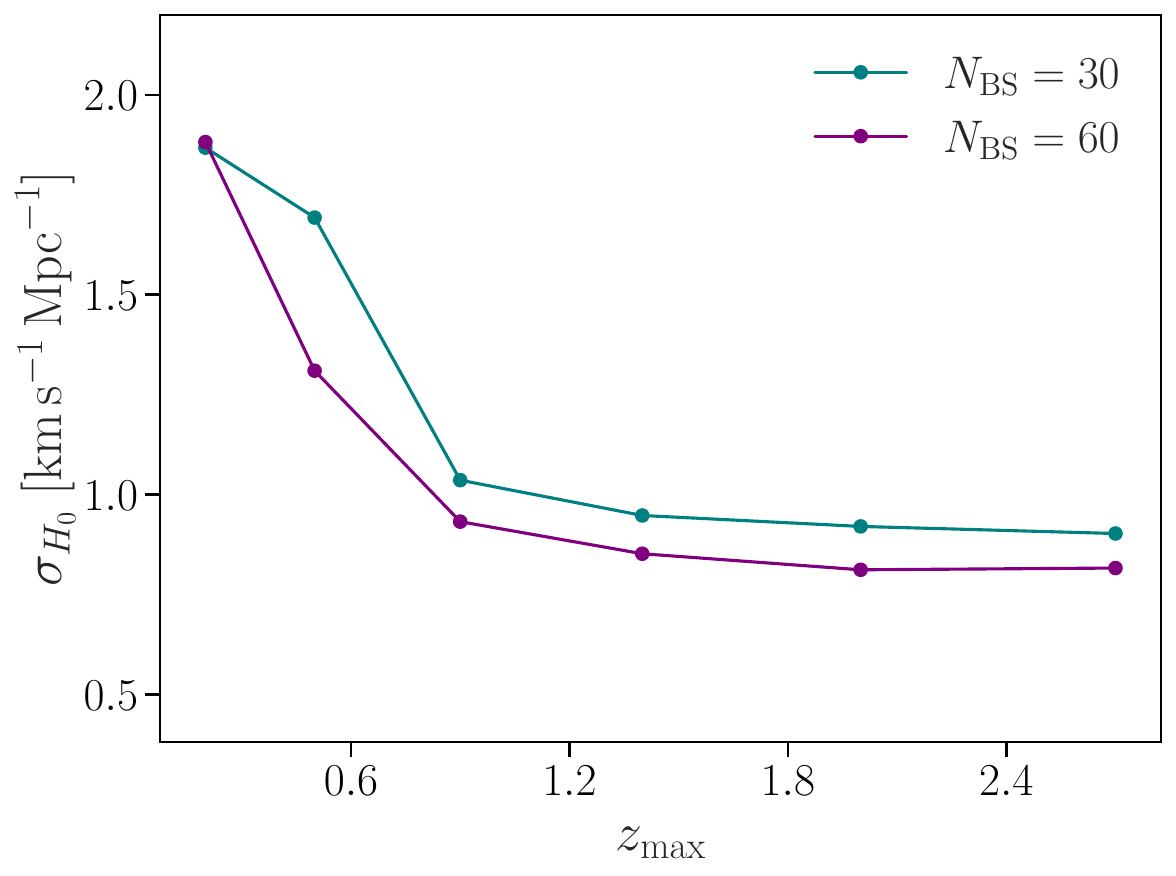}\
\textbf{(a)}
\end{minipage}
\hfill
\begin{minipage}{0.49\linewidth}
\centering
\includegraphics[width=\linewidth]{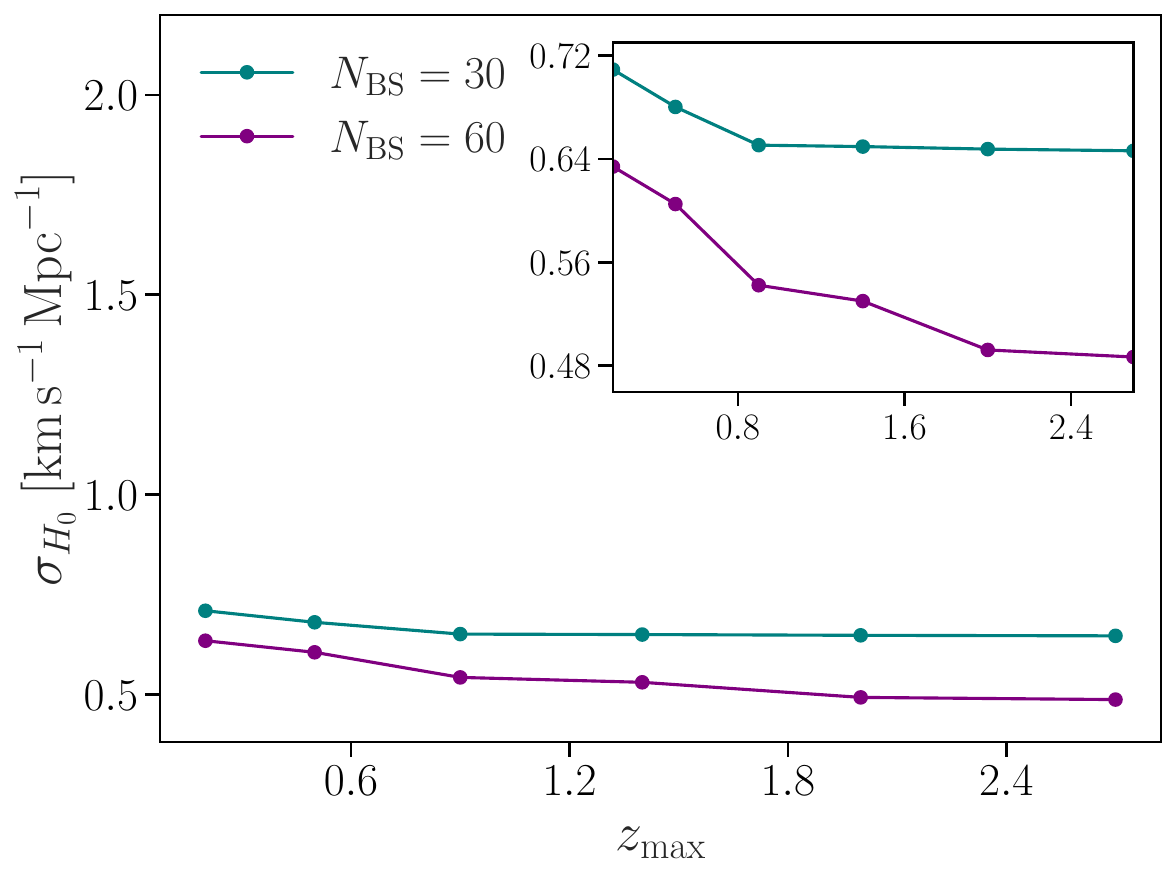}\
\textbf{(b)}
\end{minipage}

\caption{Marginalised uncertainty on $H_0$ as a function of the maximum catalogue redshift threshold $z_{\max}$, for catalogues containing $N=30$ and $N=60$ events. Each point includes all events with $z\leq z_{\max}$. Panel (a) shows the results obtained when only BS data are analysed, while panel (b) shows the corresponding results when complementary BAO information is included.}
\label{fig:H0_cumulative_comparison}
\end{figure}

We repeat the analysis cumulatively: instead of independent redshift bins, we include all events below a varying maximum redshift $z_{\max}$ and track how the marginalised uncertainty $\sigma_{H_0}$ evolves as the catalogue is extended outward in redshift. This offers an independent cross-check of the binned result above.
The results for the $N=30$ and $N=60$ catalogues are shown in Fig.~\ref{fig:H0_cumulative_comparison}. When only BS data are considered, as shown in panel (a), $\sigma_{H_0}$ decreases rapidly as the lowest-redshift events are progressively included and then largely saturates by $z_{\max}\simeq 1$. This does not imply that higher-redshift bright sirens carry no cosmological information, but rather that their additional contribution to the marginalised constraint on $H_0$ is much smaller than that already provided by the lower-redshift events.

\subsection{Impact of the number of events}
Taken together, the binned and cumulative analyses show that the lower redshift portion of the catalogue carries most of the standalone information on $H_0$, while at high redshift, luminosity-distance uncertainties are larger and the inference of $H_0$ becomes more degenerate with $\Omega_m$. For this reason, we restrict the mock catalogues used in the following BS-only forecast to $0.1 < z \leq 1$. This choice is intended to retain only the events that contribute more efficiently to the marginalised $H_0$ constraint. \cdl{For instance, for the $N_{\rm BS}=60$ catalogue, restricting to $z_{\max}\simeq 1$ yields $\sigma_{H_0}=0.93\,{\rm km\,s^{-1}\,Mpc^{-1}}$, while extending the catalogue up to $z_{\max}=2.7$ only reduces it to $\sigma_{H_0}=0.9\,{\rm km\,s^{-1}\,Mpc^{-1}}$, an improvement of about $3\%$ despite the additional high-redshift events.} \cdl{We note that this redshift cut is not intended as a way of discarding higher-redshift events in an analysis of real data or artificially improving the results. It is simply intended as a localised forecast of how many bright sirens are required to reach a given precision in the redshift range containing most of the information relevant to constrain $H_0$. Nevertheless, the higher-redshift sirens would also be considered in an analysis made with a real catalogue.}

\begin{figure}
\centering

\begin{minipage}{0.49\linewidth}
\centering
\includegraphics[width=\linewidth]{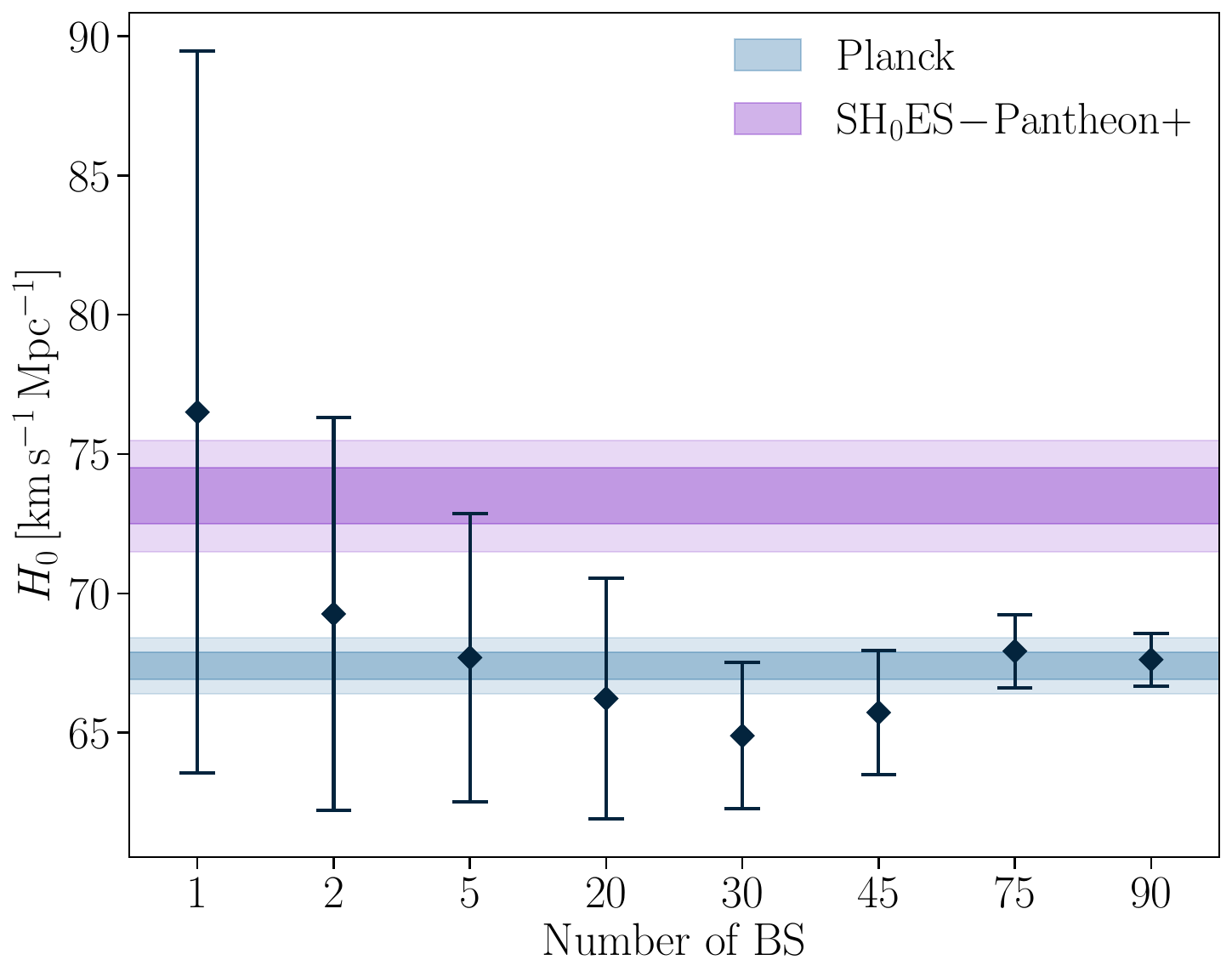}\
\textbf{(a)}
\end{minipage}
\hfill
\begin{minipage}{0.49\linewidth}
\centering
\includegraphics[width=\linewidth]{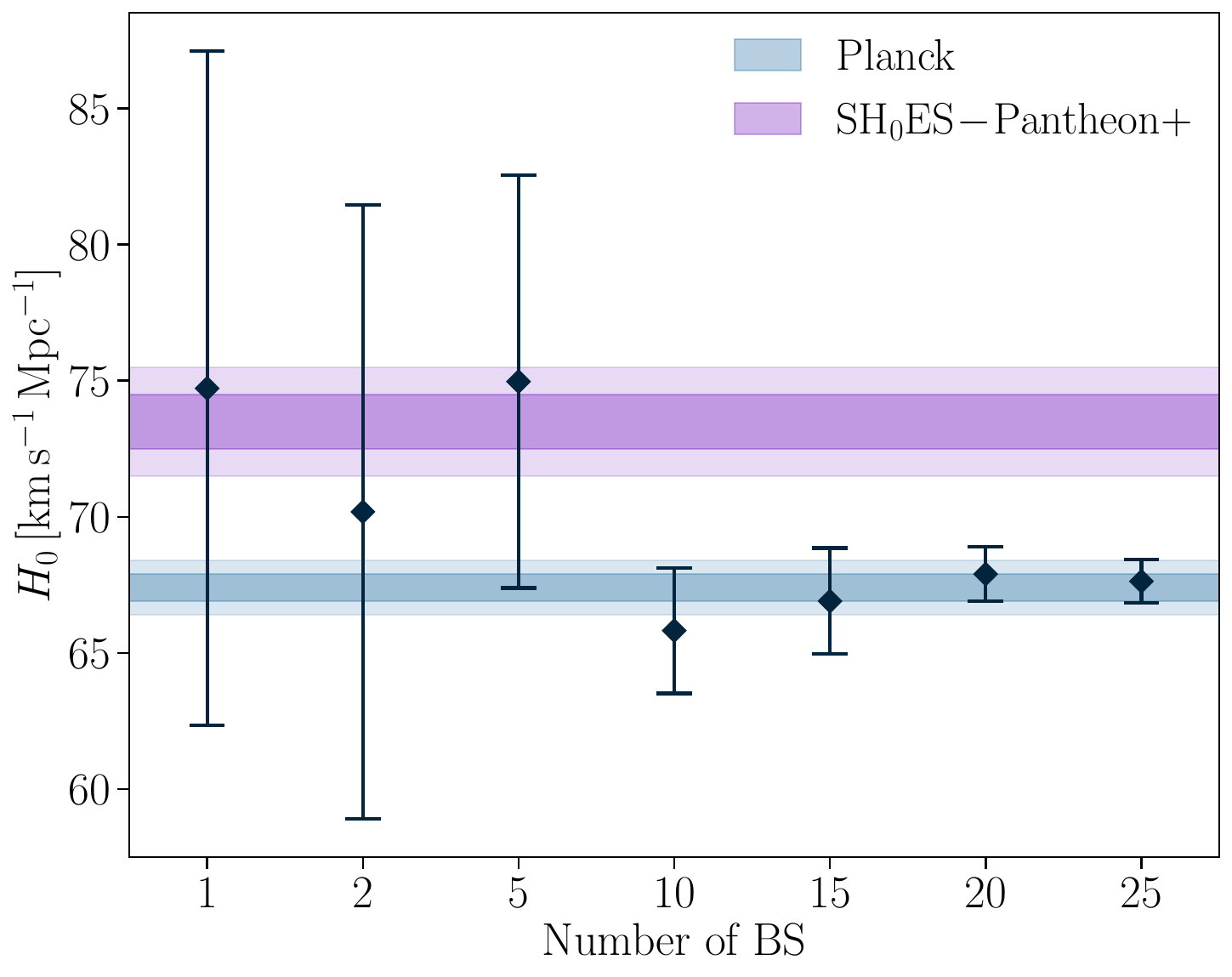}\
\textbf{(b)}
\end{minipage}

\caption{Constraints on $H_0$ as a function of the total number of bright sirens, including GW170817. Panel (a) uses bright siren catalogues restricted to the redshift range $0.1<z\leq 1.0$, with both $H_0$ and $\Omega_m$ being sampled. Panel (b) combines catalogues extending to $z\leq 3$ with external information from the SKAO BAO mock catalogue, while varying $H_0$, $\Omega_m$ and $r_d$. The purple bands show the S$H_0$ES-Pantheon$+$ constraint, with the darker and lighter regions corresponding to the $1\sigma$ and $2\sigma$ intervals, respectively. The blue bands show the corresponding \textit{Planck} reference intervals.}
\label{fig:main_H0_BS_comparison}
\end{figure}

These catalogues are generated following the same procedure adopted for the master catalogue described in Section~\ref{sec:data_bs}, filtering the events according to the redshift range of interest. Our aim is to determine how many BSs are required, within this most informative redshift range, to achieve uncertainties comparable to those of current $H_0$ measurements, focusing in particular on $\sigma_{H_0}\sim1\,{\rm km\,s^{-1}\,Mpc^{-1}}$ and $\sigma_{H_0}\sim2\,{\rm km\,s^{-1}\,Mpc^{-1}}$. The resulting constraints on $H_0$ as a function of the number of events are shown in Fig.~\ref{fig:main_H0_BS_comparison}.

The analysis shows a clear improvement in the precision of $H_0$ as the number of low-redshift BSs increases. The GW170817-only case yields a very broad posterior, which is insufficient to discriminate between the different reference values of $H_0$ \citep{Abbott2017H0}. The uncertainty already begins to decrease noticeably when only a few additional events are included, although these small catalogues still do not provide a competitive constraint. The improvement becomes progressively more pronounced as the catalogue grows, with the posterior width continuing to shrink across the different values of $N_{\rm BS}$ considered.
Note that, because the mock catalogues are generated using the \textit{Planck} best-fit value of $H_0$, the posterior means fluctuate around the fiducial \textit{Planck} value. The relevant forecast quantity is therefore the width of the posterior, rather than its overlap with either the \textit{Planck} or the S$H_0$ES-Pantheon$+$ value. In the redshift range $0.1<z\leq1.0$, we find that $\sim90$ and $\sim45$ bright sirens are required for the uncertainty to reach $\sigma_{H_0}\sim1\,{\rm km\,s^{-1}\,Mpc^{-1}}$ and $\sigma_{H_0}\sim2\,{\rm km\,s^{-1}\,Mpc^{-1}}$, respectively.

Thus, although low-redshift bright sirens provide a clean and independent measurement of $H_0$, a catalogue substantially larger than the single event currently available is required for a BS-only analysis to reach the precision of present electromagnetic and CMB-based constraints.

\clearpage

\subsection{Breaking the $H_0-\Omega_m$ degeneracy}
\label{sec:results_omegam}

\begin{figure}
\centering
\includegraphics[width=0.75\linewidth]{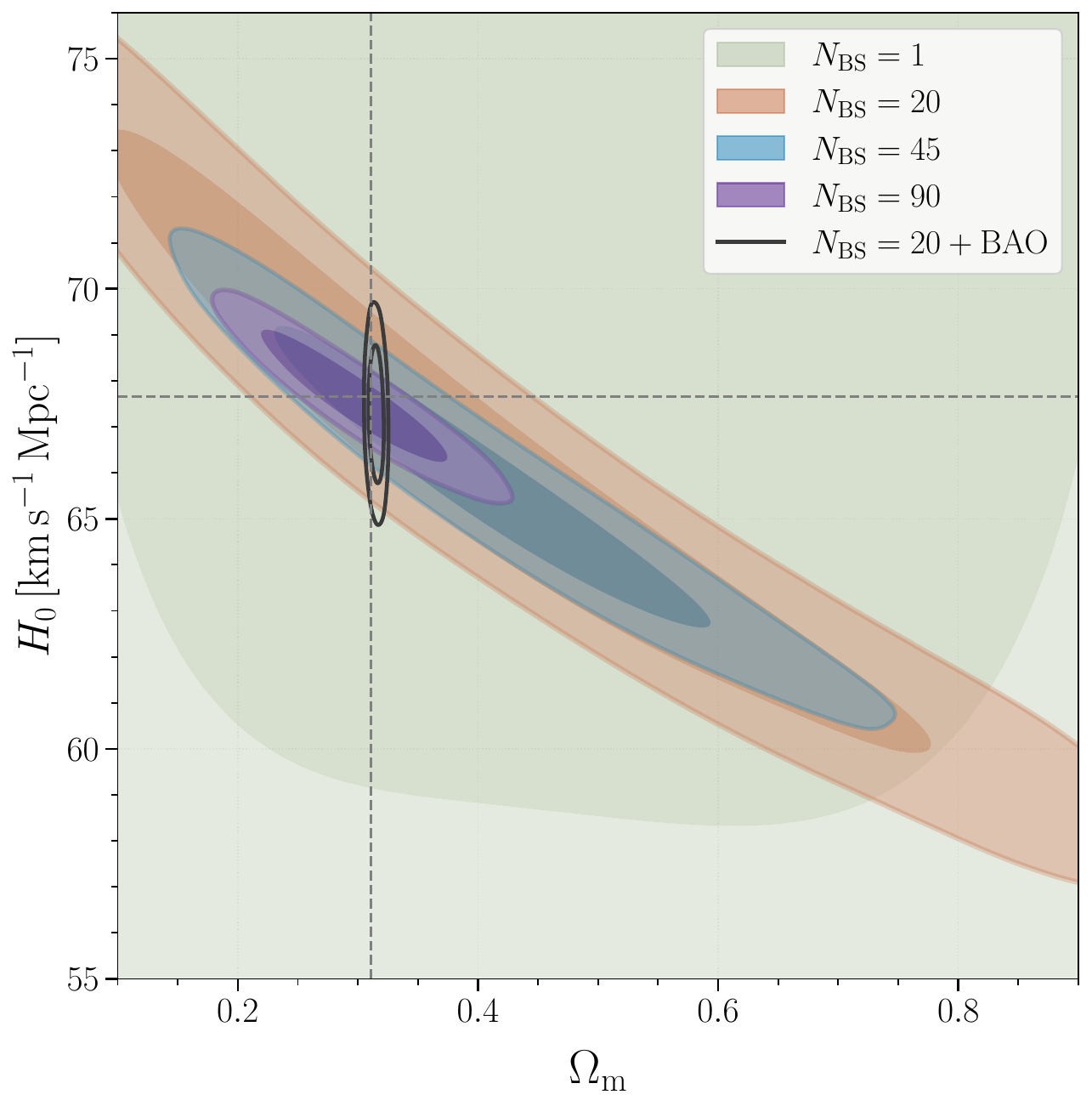}
\caption{Two-dimensional constraints in the $H_0$--$\Omega_m$ plane obtained from BS catalogues containing $N_{\rm BS}=1,20,45,$ and $90$ events, together with the constraint from the $N_{\rm BS}=20$ catalogue combined with BAO measurements. The darker and lighter coloured contours refer to the $1\sigma$ and $2\sigma$ intervals, respectively. The $N_{\rm BS}=1$ case corresponds to GW170817, while the remaining catalogues are constructed by progressively adding simulated events to it, as described in Section~\ref{sec:data_bs}.}
\label{fig:BS_H0_omegam}
\end{figure}

The degeneracy between $\Omega_m$ and $H_0$ is illustrated in Fig.~\ref{fig:BS_H0_omegam}. In a flat $\Lambda$CDM cosmology, the luminosity distance depends both on the overall distance scale set by $H_0$ and the shape of the expansion history determined by $\Omega_m$. This degeneracy is weak for sufficiently nearby sources but becomes increasingly important as the catalogue is extended to higher redshifts. This is where a constraint on $\Omega_m$ from, e.g., BAO data can help: by breaking the degeneracy, it significantly improves the constraining power of high-redshift BSs on $H_0$.

We first repeat the binned-redshift analysis after including the SKAO BAO mock. As shown in panel (b) of Fig.~\ref{fig:variance_zbins_comparison}, complementary BAO information reduces the uncertainty on $H_0$ at all redshifts, with the largest relative gain in the intermediate and high-redshift bins: once BAO data restrict the background expansion history, the same bright sirens retain substantially more information on $H_0$ than in the BS-only case. Concretely, the informative redshift range extends to $z\sim 1.5$, compared with $z\lesssim 1$ when BS are analysed alone. The variance still increases for the most distant events, whose distance uncertainties are intrinsically larger, but it remains comparable to the uncertainty found at $z\sim 1$ in the BS-only analysis.

The corresponding cumulative analysis, shown in panel (b) of Fig.~\ref{fig:H0_cumulative_comparison}, confirms this picture. Unlike the BS-only case, $\sigma_{H_0}$ shows no pronounced saturation at $z_{\max}\sim 1$: it continues to decrease as intermediate-redshift events are added, with a gain that shrinks only gradually, so that sources out to $z_{\max}\sim 2$ still contribute visibly. 
This behaviour is also reflected in Fig.~\ref{fig:BS_H0_omegam}, where the black $N_{\rm BS}=20+\mathrm{BAO}$ contours are substantially narrower than the corresponding BS-only contours, demonstrating that the inclusion of BAO information significantly reduces the $H_0$--$\Omega_m$ degeneracy.
Complementary BAO information therefore does more than tighten the final constraint, it also reshapes the redshift range over which the BS catalogue contributes efficiently to the $H_0$ constraint.

Motivated by this result, for the BS-BAO analysis we extend the mock catalogue up to $z_{\max}=3.0$. 
The resulting constraints are shown in panel (b) of Fig.~\ref{fig:main_H0_BS_comparison}. 

First, we note that the substantial gain in precision between the $N=5$ and $N=10$ BS catalogues is partly driven by the specific realisation. Since the sub-catalogues are constructed by randomly adding individual events, the properties of the few sources introduced in this interval—such as a particularly well-localised event or relatively high redshift events—can have an outsized impact when the sample is still small. The overall improvement is nevertheless physically expected, since the $H_0$--$\Omega_m$ degeneracy can be efficiently reduced once the BS contour becomes sufficiently narrow to combine effectively with the BAO constraint. However, the sharp knee observed at $N=10$ should not be over-interpreted as a fixed, reproducible number.

We find that approximately $N\sim20$ and $N\sim15$ events are sufficient to reach a precision of $\sigma_{H_0} \sim 1\,{\rm km\,s^{-1}\,Mpc^{-1}}$ and $\sigma_{H_0} \sim 2\,{\rm km\,s^{-1}\,Mpc^{-1}}$, respectively.
 \cdl{To assess how representative this realisation is, in Sec.~\ref{sec:realisations} we repeat the BS+BAO analysis on 1000 realisations of the sub-catalogues. We find that $\sigma_{H_0}\le2\,{\rm km\,s^{-1}\,Mpc^{-1}}$ is reached with $N_{\rm BS}\simeq14$ for the median realisation, confirming the result above, while $\sigma_{H_0}\le1\,{\rm km\,s^{-1}\,Mpc^{-1}}$ requires $N_{\rm BS}\simeq46$ for the median realisation. The realisation used here is therefore a favourable one.}
 
The comparison highlights the central result of this work: the inclusion of information from BAO on $\Omega_m$ enables a substantial improvement with respect to the BS-only case, where we found that about $N\sim90$ events are required to reach the same level of precision.

\section{Robustness to the catalogue realisation}
\label{sec:realisations}

In Sec.~\ref{sec:results} we showed that the constraint on $H_0$ depends on the redshift distribution of the bright sirens. When BSs are analysed alone, the information on $H_0$ comes almost entirely from low-redshift events: at intermediate redshift the luminosity distance also depends on $\Omega_m$, and the strong $H_0$--$\Omega_m$ degeneracy prevents these events from adding much to the $H_0$ constraint. When BSs are combined with BAO, low-redshift events remain the most informative, but intermediate-redshift events also become important. BAO constrain $\Omega_m$ and thereby break the $H_0$--$\Omega_m$ degeneracy, so the distances measured by intermediate-redshift BSs translate into significant additional information on $H_0$.

The results of Sec.~\ref{sec:results} are, however, obtained from a single realisation of the mock. The sub-catalogues are built by randomly drawing BS events from the master catalogue described in Sec.~\ref{sec:data_bs}, and adding events progressively as $N_{\rm BS}$ grows. A single realisation is what we will actually observe. Nonetheless, since the redshift distribution of the events has a strong impact on $\sigma_{H_0}$, the particular catalogue used in the analysis presented in Sec.~\ref{sec:results}  could be a favourable (or unfavourable) draw. \ett{Catalogue fluctuations have also been considered in previous ET bright siren forecasts \citep[e.g.][]{Canevarolo:2023dkh}.} Here we quantify how much $\sigma_{H_0}$ varies between realisations, and how many BSs are required to reach a given precision in a statistical sense.

\subsection{Realisations and analysis}

We repeat the BS+BAO analysis of Sec.~\ref{sec:results_omegam} for $N_{\rm BS}\in\{20,25,30,35,40,45\}$, using up to 1000 independent realisations of the master mock catalogue. As explained in Sec.~\ref{sec:data_bs}, each sub-catalogue contains GW170817 plus $N_{\rm BS}-1$ simulated events. As in the previous analysis, the sub-catalogues built from a given realisation are nested: we first draw 19 simulated events from that realisation to build the $N_{\rm BS}=20$ sub-catalogue, and then add 5 new events at each step to obtain the sub-catalogues with $N_{\rm BS}=25, 30, \dots, 45$. In this way, within a given realisation, the change in the constraints with $N_{\rm BS}$ is due only to the newly added events, while the scatter across realisations captures the variability due to the specific set of observed events.

\begin{figure*}
    \centering
    \includegraphics[width=\textwidth]{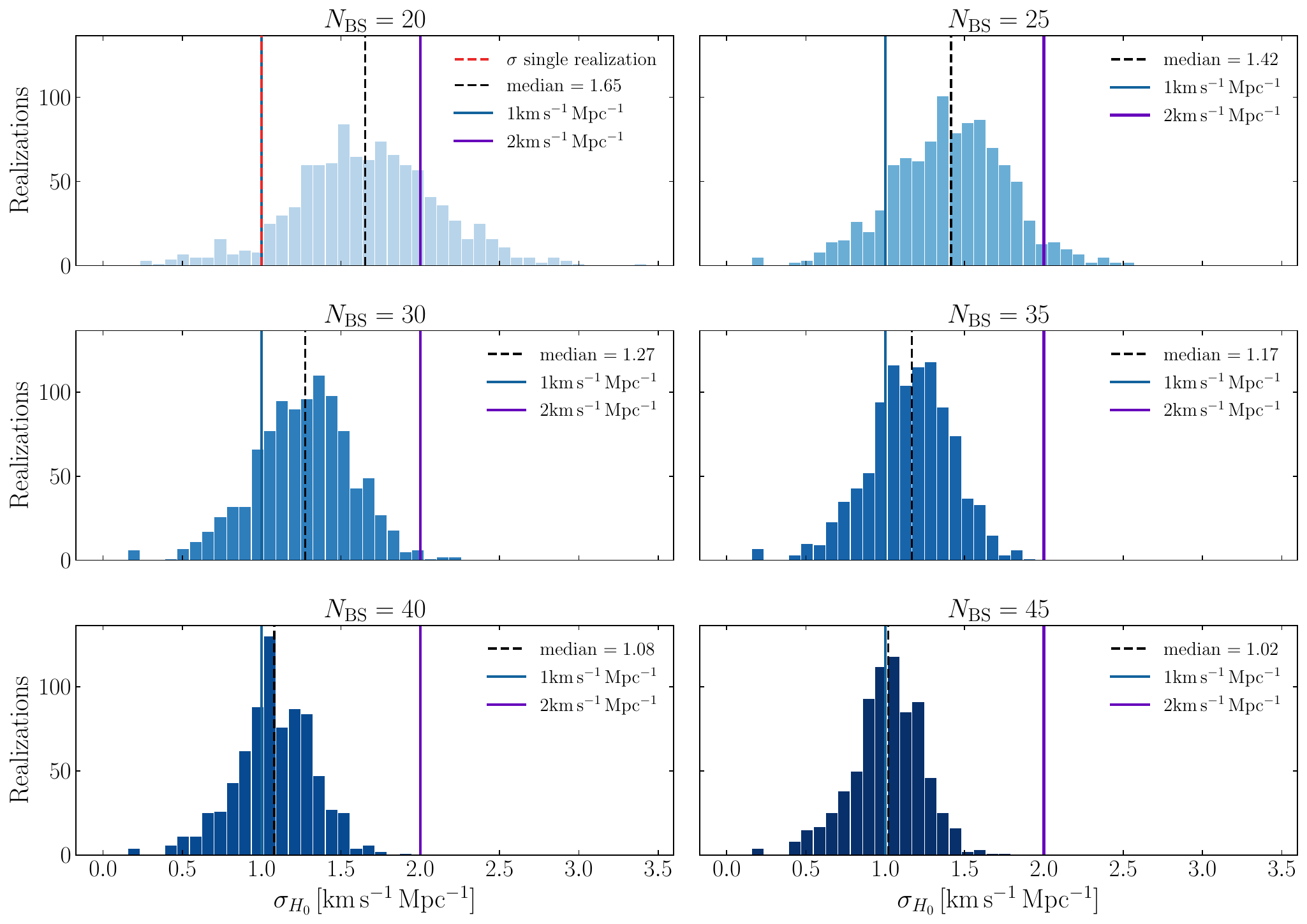}
    \caption{Distribution of $\sigma_{H_0}$ over the realisations for $N_{\rm BS}=20$--$45$, from the BS+BAO analysis. The black dashed lines show the median of each distribution. The blue and purple lines mark the target precisions of $1$ and $2\,{\rm km\,s^{-1}\,Mpc^{-1}}$. In the $N_{\rm BS}=20$ panel, the red dashed line shows the value obtained from the single realisation used in Sec.~\ref{sec:results}.}
    \label{fig:sigma_distribution}
\end{figure*}

The distribution of $\sigma_{H_0}$ for the six values of $N_{\rm BS}$ is shown in Fig.~\ref{fig:sigma_distribution}, together with its median and the two target precisions, $1$ and $2\,{\rm km\,s^{-1}\,Mpc^{-1}}$. As expected, the distribution shifts towards smaller values and becomes narrower as $N_{\rm BS}$ increases. The median goes from $1.65\,{\rm km\,s^{-1}\,Mpc^{-1}}$ for $N_{\rm BS}=20$ to $1.02\,{\rm km\,s^{-1}\,Mpc^{-1}}$ for $N_{\rm BS}=45$.
 
For $N_{\rm BS}=20$ we also show the value obtained in the single realisation analysis, $\sigma_{H_0}\simeq1\,{\rm km\,s^{-1}\,Mpc^{-1}}$. It lies in the lower tail of the distribution: it is inside the central 95\% interval but below the 16th percentile, and only about 5--10\% of the realisations reach it. The main-text catalogue is therefore a favourable, but not exceptional, realisation. In a statistical sense, the median realisation requires $N_{\rm BS}\simeq46$ to reach $\sigma_{H_0}=1\,{\rm km\,s^{-1}\,Mpc^{-1}}$, while $\sigma_{H_0}=2\,{\rm km\,s^{-1}\,Mpc^{-1}}$ is reached by most realisations already for $N_{\rm BS}=20$ (see Table~\ref{tab:Nreq}).

\subsection{Modelling $\sigma_{H_0}$ as a function of $N_{\rm BS}$}
To predict the number of sirens needed beyond the simulated range, we model how the distribution of $\sigma_{H_0}$ evolves with $N_{\rm BS}$. At each $N_{\rm BS}$ we compute five percentiles $\sigma_q$ of the distribution: the median ($q=50\%$), the edges of the 68\% interval ($q=16\%,84\%$) and the edges of the 95\% interval ($q=2.5\%,97.5\%$). Their statistical uncertainties $\delta\sigma_q$ are obtained with the resampling procedure described at the end of this section. Each percentile is then fitted with a power law,
\begin{equation}
    \sigma_q(N_{\rm BS}) = \sigma_{0,q}\left(\frac{N_{\rm BS}}{45}\right)^{-\alpha_q}\,,
    \label{eq:powerlaw}
\end{equation}
where $\alpha_q$ sets how fast $\sigma_q$ decreases as more sirens are added, and $\sigma_{0,q}$ is the value of $\sigma_q$ at $N_{\rm BS}=45$.
For identical, independent events with fixed $\Omega_m$ one would expect $\alpha=1/2$. For the median we find
\begin{equation}
    \sigma_{\rm med}(N_{\rm BS}) = (1.01\pm0.02)\,{\rm km\,s^{-1}\,Mpc^{-1}}\left(\frac{N_{\rm BS}}{45}\right)^{-0.59\pm0.03}\,.
\end{equation}
This means that the typical realisation reaches $\sigma_{H_0}\simeq1\,{\rm km\,s^{-1}\,Mpc^{-1}}$ with about 46 sirens.
%The uncertainties on the amplitude and on the slope are correlated, because both are estimated from the same points. For the other percentiles, $\alpha_q$ ranges from $0.45\pm0.13$ for the 2.5th percentile to $0.71\pm0.03$ for the 97.5th. The power law reproduces the measured percentiles to better than 5\% at every $N_{\rm BS}$, and to about 1\% for the median.

\ett{The effective scaling $\alpha_{\rm med}=0.59\pm0.03$ is slightly steeper than the $N_{\rm BS}^{-1/2}$ behaviour expected for identical independent measurements \cite{Taylor:2012db}. This is consistent with the fact that each new event improves the constraint on $H_0$ directly, accompanied by a progressive reduction of the $H_0-\Omega_m$ degeneracy when the sirens are combined with BAO, as described in Sec.~\ref{sec:results_omegam}.}

\begin{figure}
    \centering
    \includegraphics[width=\columnwidth]{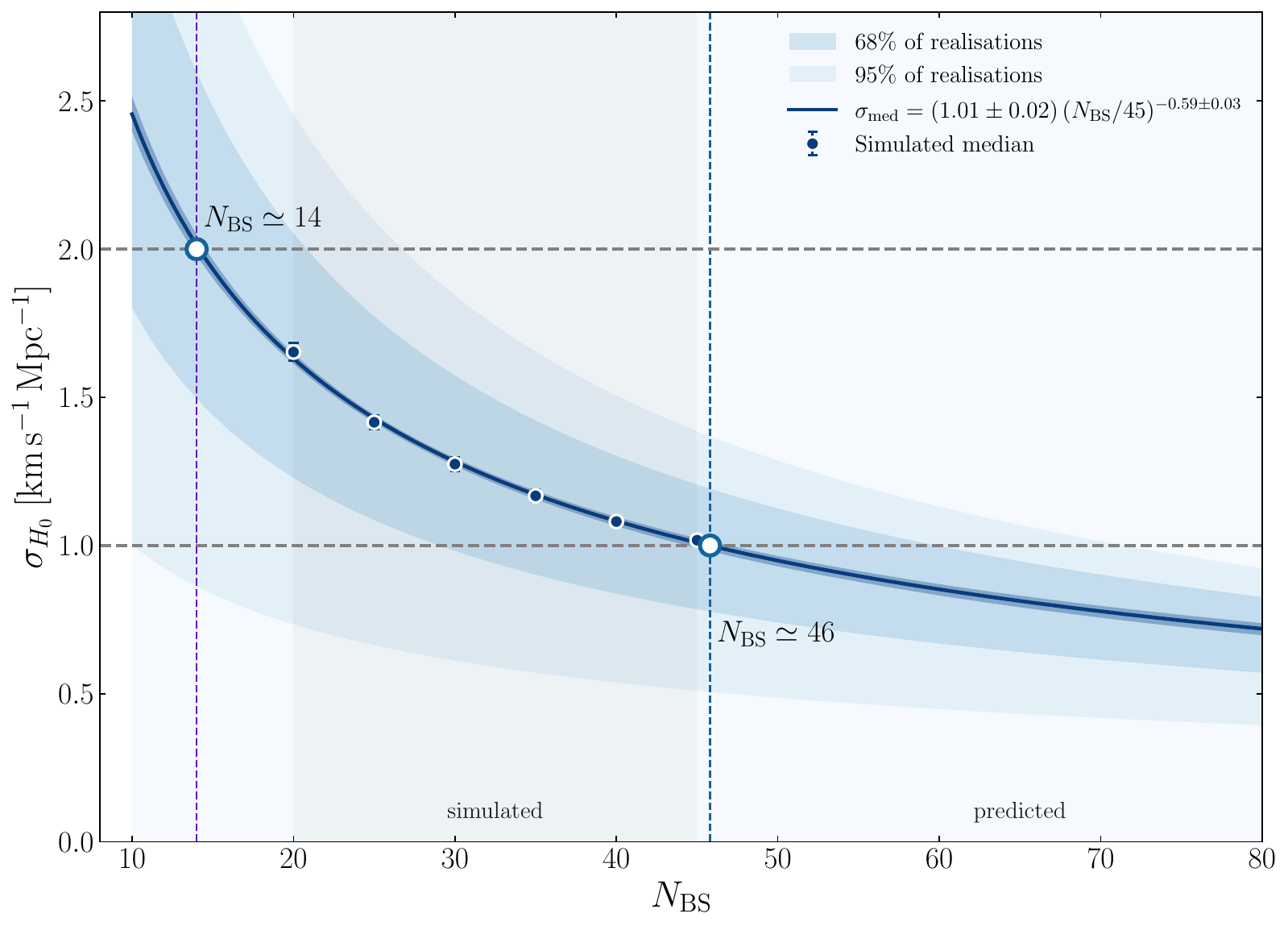}
\caption{Median $\sigma_{H_0}$ as a function of $N_{\rm BS}$ (points) and the power-law fit to it (solid line). The shaded bands are power-law fits to the 16th--84th and 2.5th--97.5th percentiles of $\sigma_{H_0}$, i.e.\ the ranges containing 68\% and 95\% of the realisations, and show the spread between catalogue realisations. The grey region marks the simulated range, $20\le N_{\rm BS}\le45$, and the light-blue regions on either side the extrapolation. The white circles mark where the median prediction reaches $\sigma_{H_0}=2$ and $1\,{\rm km\,s^{-1}\,Mpc^{-1}}$ (dashed horizontal lines), at $N_{\rm BS}\simeq14$ and $\simeq46$, respectively (dashed vertical lines). \vspace{3 mm}}
    \label{fig:sigma_prediction}
\end{figure}

Figure~\ref{fig:sigma_prediction} shows the predicted $\sigma_{\rm med}$ as a function of $N_{\rm BS}$, together with two different kinds of uncertainty. The wide shaded regions show how much $\sigma_{H_0}$ varies between realisations: at a given $N_{\rm BS}$, 68\% and 95\% of the realisations fall inside them. This spread depends on which events are observed, so it does not shrink if more realisations are simulated. The error bars on the points and the narrow band around the median line show instead how well we know the median itself. This uncertainty comes only from having a finite number of realisations, so it would shrink if more were simulated. Both are obtained with the resampling procedure described below.

A fraction $q$ of the realisations reaches a target precision $\sigma_\star$ once the $q$-th percentile falls below it. Inverting Eq.~\eqref{eq:powerlaw}, the number of events required is
\begin{equation}
    N_{\rm req}(q) = 45\left(\frac{\sigma_{0,q}}{\sigma_\star}\right)^{1/\alpha_q}\,.
\end{equation}
We quote $q=50\%$ (typical realisation), $84\%$ and $97.5\%$, i.e.\ the upper edges of the central 68\% and 95\% intervals of $\sigma_{H_0}$, for $\sigma_\star=1$ and $2\,{\rm km\,s^{-1}\,Mpc^{-1}}$, which correspond to twice the $1\sigma$ uncertainties of Planck and SH0ES, respectively.

We consider two sources of uncertainty on $N_{\rm req}$ and on the fitted parameters.

Since we have a finite number of realisations, the percentiles of the $\sigma_{H_0}$ distribution, and every quantity derived from them, are themselves affected by a statistical uncertainty: a different set of realisations would give slightly different values. We estimate this uncertainty directly from the data by resampling. We build $N_{\rm boot}=1000$ new sets of realisations, each obtained by drawing at random, with replacement, from the original set. Each resampled set therefore has the same size as the original one, but some realisations appear more than once and others not at all. For example, if the original set contained only five realisations, $\{r_1,r_2,r_3,r_4,r_5\}$, a resampled set could be $\{r_2,r_2,r_4,r_5,r_1\}$: $r_2$ is drawn twice and $r_3$ is not drawn. On each resampled set we recompute the percentiles of $\sigma_{H_0}$ at every $N_{\rm BS}$, the power-law fit of Eq.~\eqref{eq:powerlaw} and the derived $N_{\rm req}$. The spread of a quantity across the 1000 resampled sets, i.e.\ its standard deviation, is taken as its statistical uncertainty. For example, the power-law fit to the median of the original set gives a slope $\alpha_{\rm med}=0.591$. Refitting each of the 1000 resampled sets gives 1000 slightly different slopes with a standard deviation of $0.026$, so we quote $\alpha_{\rm med}=0.59\pm0.03$. In the same way, the original set gives $N_{\rm req}=45.8$ for the median realisation to reach $\sigma_{H_0}=1\,{\rm km\,s^{-1}\,Mpc^{-1}}$, and the scatter of the values obtained from the resampled sets gives a statistical error of $1.4$. This procedure provides the uncertainties $\delta\sigma_q$ used as weights in the fit, the errors on $\alpha_q$, the statistical error on $N_{\rm req}$, and the 68\% band around the median prediction in Fig.~\ref{fig:sigma_prediction}.

In addition, the predicted $N_{\rm req}$ also depends on the function we choose to fit. To estimate this effect, we repeat the calculation with two alternative fits: (i) the same power law, fitted only to the points $N_{\rm BS}=25$--$35$; (ii) a model in which each new event adds the same amount of information on $H_0$. Since the information is proportional to $1/\sigma^2$, the second model reads $1/\sigma_q^2=a_q+b_q\,N_{\rm BS}$, where $a_q$ is the information already provided by BAO and GW170817 and $b_q$ is the information added by each new siren. For large $N_{\rm BS}$ this gives the familiar $\sigma\propto N_{\rm BS}^{-1/2}$. The systematic error is the largest difference between the $N_{\rm req}$ obtained from these two fits and the baseline value. The statistical and systematic uncertainties are added in quadrature. The results are given in Table~\ref{tab:Nreq} and shown in Fig.~\ref{fig:sigma_prediction}.

 \renewcommand{\arraystretch}{1.5}

\begin{table}
    \centering
    \caption{Number of bright sirens (BS+BAO) needed for a fraction $p$ of the realisations to reach $\sigma_{H_0}\le\sigma_\star$. The quoted uncertainties combine in quadrature the statistical error, due to the finite number of realisations, and the systematic error associated with the choice of the fitting function.}
    \label{tab:Nreq}
    \begin{tabular}{lccc}
        \hline\hline
        $\sigma_\star\ [{\rm km\,s^{-1}\,Mpc^{-1}}]$ & $q=50\%$ & $q=84\%$ & $q=97.5\%$ \\
        \hline
        $1$ & $46\pm2$ & $60\pm3$ & $72\pm9$ \\
        $2$ & $14\pm1$ & $21\pm1$ & $27\pm1$ \\
        \hline\hline
    \end{tabular}
\end{table}

For $\sigma_\star=2\,{\rm km\,s^{-1}\,Mpc^{-1}}$, about 14 BSs are sufficient for the typical realisation, about 21 for 84\% and about 27 for 97.5\% of the realisations, consistent with the single-realisation analysis. For $\sigma_\star=1\,{\rm km\,s^{-1}\,Mpc^{-1}}$, the typical realisation requires $N_{\rm BS}=46\pm2$, while $N_{\rm BS}\simeq60$ and $\simeq72$ are needed for 84\% and 97.5\% of the realisations (Table~\ref{tab:Nreq}). Of these values, $N_{\rm BS}\simeq27$ lies within the simulated range $20\le N_{\rm BS}\le45$ and $21$ and $46$ at its edges, while $14$, $60$ and $72$ rely on extrapolating the power-law fit and carry a larger systematic uncertainty, which grows with the distance from the simulated range. The value $N_{\rm BS}\simeq20$ found in the single-mock analysis for $\sigma_{H_0}\simeq1\,{\rm km\,s^{-1}\,Mpc^{-1}}$ should therefore be read as achievable for a favourable redshift distribution of the detected events, rather than as the typical requirement.

\section{Conclusions}\label{sec:conclusions}

In this work, we investigated the observational requirements for future bright siren catalogues to constrain the Hubble constant with a precision comparable to current electromagnetic and CMB measurements. In particular, we studied how the constraining power depends on the number of detected bright sirens, their redshift distribution, and the inclusion of external cosmological information.

We first analysed the impact of the number and redshift distribution of the events by constructing several sub-catalogues from a common master catalogue, each containing a different number of bright sirens. We divided the events into six redshift bins and evaluated the marginalised variance of the $H_0$ posterior in each bin. 

The binned analysis showed that the variance of $H_0$ increases significantly at $z\gtrsim 1$. Consequently, adding events at these redshifts produces only a limited improvement in the marginalised constraint on $H_0$. This behaviour was confirmed by the cumulative analysis, in which events were progressively included up to increasing values of the maximum redshift. We found that the uncertainty decreases rapidly when low-redshift events are added, but becomes nearly constant once sources beyond $z\simeq 1$ are included.
This behaviour reflects two compounding effects: the increasing sensitivity of the luminosity distance to $\Omega_m$ at higher redshift, and the growing uncertainty on the distance measurement itself for more distant sources. Since bright sirens alone cannot efficiently break the resulting $H_0$--$\Omega_m$ degeneracy, high-redshift events carry comparatively little information on $H_0$ once $\Omega_m$ is left free to vary.

Motivated by these findings, we restricted the final bright siren-only catalogue to the range $0.1<z\leq 1.0$ and constructed sub-catalogues with progressively increasing numbers of events. We found that approximately $N\sim45$ and $N\sim90$ bright sirens are required to reach a precision of $\sigma_{H_0} \sim 2\,{\rm km\,s^{-1}\,Mpc^{-1}}$ and $\sigma_{H_0} \sim 1\,{\rm km\,s^{-1}\,Mpc^{-1}}$, respectively.

We then repeated the analysis by including external information from the SKAO-like BAO mock catalogue. The binned and cumulative analyses showed that, once the $H_0$--$\Omega_m$ degeneracy is reduced, bright sirens at intermediate and higher redshifts  provide substantially more information on $H_0$: the uncertainty continues to decrease out to $z\sim 2$, rather than saturating at $z\simeq 1$ as in the BS-only case, although the most distant events remain limited by their larger luminosity-distance uncertainties.
\ett{For the particular catalogue realisation analysed in Sec.~\ref{sec:results}, approximately $N\sim15$ and $N\sim20$ events are sufficient to reach $\sigma_{H_0}\sim2,{\rm km,s^{-1},Mpc^{-1}}$ and $\sigma_{H_0}\sim1,{\rm km,s^{-1},Mpc^{-1}}$, respectively.}

Since the constraining power depends strongly on the redshift distribution of the detected events, we tested how representative these numbers are by repeating the BS+BAO analysis over 1000 catalogue realisations with $N_{\rm BS}=20$--$45$. \ett{We find that the $N_{\rm BS}\simeq20$ result for $\sigma_{H_0}\simeq1\,{\rm km\,s^{-1}\,Mpc^{-1}}$ corresponds to a favourable realisation rather than the typical expectation. For the median realisation, $N_{\rm BS}\simeq14$ (extrapolated) and $N_{\rm BS}\simeq46$ are required to reach $\sigma_{H_0}=2$ and $1\,{\rm km\,s^{-1}\,Mpc^{-1}}$, respectively, while $N_{\rm BS}\simeq21$ and $\simeq60$ are required for 84\% of the realisations to reach the same precision. The median uncertainty scales as $\sigma_{H_0}\propto N_{\rm BS}^{-0.59\pm0.03}$, somewhat faster than the $N_{\rm BS}^{-1/2}$ expected for identical independent measurements, which we attribute to the progressive breaking of the $H_0$--$\Omega_m$ degeneracy when bright sirens are combined with BAO.}

Overall, our results show that bright sirens are a promising independent probe of $H_0$, in particular in the context of the Hubble tension. However, their constraining power is driven not only by the total number of detected events, but also by their redshift distribution and by degeneracies with the other cosmological parameters. Combining bright sirens with complementary probes such as BAO is therefore particularly valuable: it extends the redshift range over which the catalogue is informative and substantially reduces the number of events needed for a competitive measurement of $H_0$.

\section*{Acknowledgments}

We thank Matteo Martinelli for useful discussions and feedback on earlier versions of this draft. 
CDL acknowledges financial support from Sapienza Università di Roma, provided through Progetti Medi 2021 (Grant No. RM12117A51D5269B). 
EMT and VP are supported by funding from the European Research Council (ERC) under the European Union’s HORIZON-ERC-2022 (grant agreement no. 101076865).
This work made use of Melodie, a computing infrastructure funded by the same project, and PLEIADI, a computing infrastructure installed and managed by INAF. 
This article is based upon work from COST Action CA21136 Addressing observational tensions in cosmology with systematics and fundamental physics (CosmoVerse) supported by COST (European Cooperation in Science and Technology).

\bibliographystyle{aasjournal}
\bibliography{biblio}

\end{document}